# Home Energy Management in Smart Grid

A. Mahmood[1], H. Fakhar[1], S. H. Ahmed[2], N. Javaid[1,*]

COMSATS Institute of Information Technology, Islamabad Pakistan.
School of Computer Science & Engineering, Kyungpook National University, Korea.

*Abstract-* **A significant amount of research has been conducted in order to make home appliances more efficient in terms of energy usage. Various techniques have been designed and implemented in order to control the power demand and supply. This paper encompasses reviews of different research works on a wide range of energy management techniques for smart homes aimed at reducing energy consumption and minimizing energy wastage. The idea of smart home is elaborated followed by a review of existing energy management methods.**

## I. INTRODUCTION

A home constituting of such appliances or devices that consume energy in an efficient manner and operate digitally is known as a smart home [1]. A network oriented home makes communication between devices in an easy manner. It enables the devices to be connected to the internet server and usually communicates wirelessly. The internet or service providers are the ones responsible for providing new services to the customers and making them easily accessible for the individuals. In some cases, the smart devices are supposed to be smart enough to observe the inhabitants residing in these homes. Latest technologies are used to build such devices that operate and communicate automatically.

The sole purpose of these smart devices is to provide its residents with a secure, active and quality environment leading to a positive impact on their lives. The issues such as environmental disruptions, insecurity, communication difficulties, health dilemmas and entertainment discrepancies are reduced to a minimum by the use of smart devices. Smart home devices are capable of monitoring internal activities of the home and use technology that enables the devices, for example, to turn on and turn off automatically. Concept of the smart home is elaborated in Figure 1. It shows different smart devices to be controlled for efficient energy management.

Rest of the paper is organized as follows. Section II is dedicated for energy regulation methods used in smart homes where as section III describes the quest for efficient energy resources and energy management using ZigBee. Conclusions are drawn in section IV.

## II. ENERGY MANAGEMENT IN SMART HOMES

Various energy management methods for smart homes are described in this section. The use of neural fuzzy network has been in use to control devices in the house. A controller based upon a neural fuzzy logic was designed in MATLAB interface and it was converted into a tool with the help of hardware and the use of internet technology [2]. However, this technique is to be used in automating home appliances and is applied in order to manage time and energy used by the appliances.

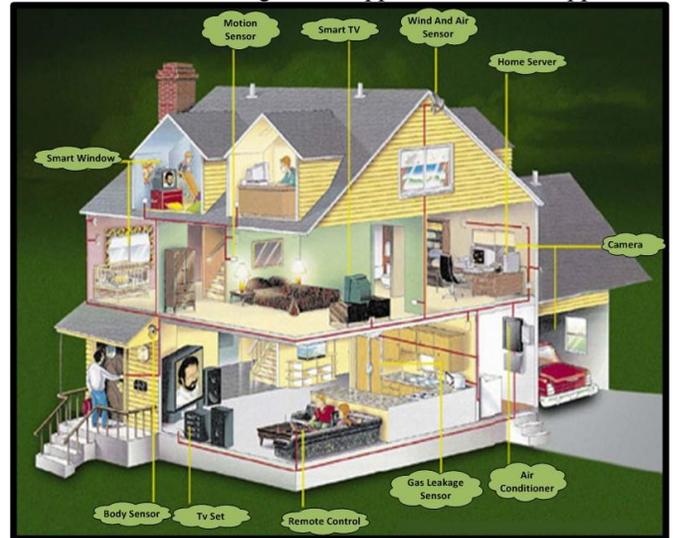

Fig. 1. Smart appliances for efficient energy management

A method to conserve energy build upon home information state has been used in [3]. In order to manage and control appliances that work on the principle of user's operational information, a network including IP, XML and JAVA is present at most of the companies dealing in appliances used at home. An experiment was carried out by simulating an environment consisting of hundred homes in which 16 different home appliances operating electrically were used. The simulation was carried out carefully and in a very professional manner so that accurate results could be achieved. The occupants of these homes were told to behave normally while the devices were being operated by using the neural fuzzy technology. The results showed that the appliances helped in reducing the energy consumption by 15.6% in contrast to those devices which were operating manually. Another system to manage energy has been presented in [4]; it consists of a simple desktop PC and some sensors in a building. The purpose was to ensure the conservation of energy that is lost by carelessness caused by users. The sensors are used in order to see how many individuals entered and exited the building.

An energy management system was developed that operated intelligently inside a small building occupied by inhabitants [5]. It basically works by observing the activities of individuals using thermal and visual aid. The sensors were used to monitor the activities by sensing the heat level of the people. By building a system that operated visually and thermally, the energy wastage was reduced to a great level. A smart home system was designed to control the energy consumption in household for demand side management DSM in [6]. The idea here is to anticipate the load or demand of energy required to run the appliances inside the home. In [7], the concept of conserving energy was taken to a whole new level with a broader limit. A management system that operated automatically was deployed in a residential area where the idea was to modify the consumption of power and limit the use of energy to a minimum level based upon past events and records. Role of the ZigBee in home energy management and its comparison with Bluetooth and Wi-Fi is presented in the subsequent section.

### III. QUEST FOR EFFICIENT ENERGY RESOURCES AND ENERGY MANAGEMENT USING ZIGBEE

Over the years, due to increase in the electricity tariff and installation of excessive home appliances, residential power consumption has increased rapidly. Increased power consumption and exhausted natural energy resources have forced the humans to find alternate energy resources. The problems such as global warming, pollution, matter decay, ozone depletion and the like have increased the use of natural resources which are not in abundance to meet the energy demands of the modern world. Moving towards finding efficient energy source will require changing the way energy is supplied as well as used [8]. The communication protocol used has to be efficient as well as continuous so that the devices being operated provide an un-interrupting service.

In most power related applications the use of wireless technology can bring profit to the end users by integrating wireless technology, output optimization etc. ZigBee is the popular wireless network used in home energy management system. It is a general communication protocol that consumes less power while being economical at the same time. Features of the Zigbee protocol are shown in Figure 2. A brief comparison of ZigBee with other technologies is presented in Table 1.

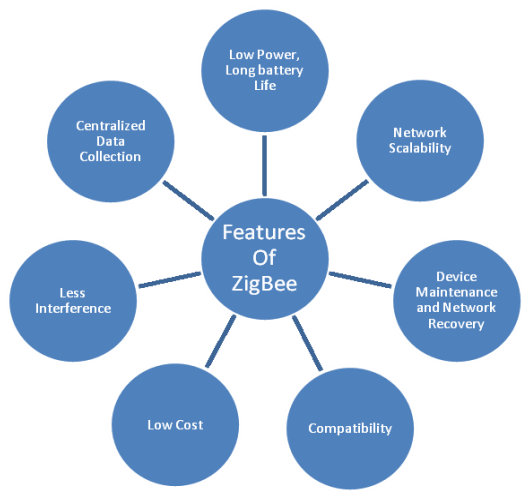

Fig. 2. Features of ZigBee

TABLE I
COMPARISON OF ZIGBEE WITH OTHER TECHNOLOGIES

| Salient | Bluetooth | Wi-Fi | ZigBee |
|---|---|---|---|
| Ease of Use | Normal | Hard | Easy |
| Prime Cost | Low | Normal | Low |
| Reliability | High | Normal | High |
| Security | 62 bit, 128 bit | SSID | 128 bit AES |
| Cost of Use | None | None | None |
| Cost of terminal unit | Low | High | Low |
| Linking time | Up to 10 s | Up to 3 s | 30 ms |
| Network Nodes | 8 | 50 | 65535 |

### IV. CONCLUSIONS

Methods for home energy management in smart grid are reviewed in this paper. Smart homes and energy management will play important role in demand side management of future smart grids. Features and role of ZigBee for smart homes have also been covered. ZigBee is compared with other wireless technologies in order to prove its suitability for home energy management.

REFERENCES

[1] V. Ricquebourg, D. Menga, D. Durand, B. Marhic, L. Delahoche, C. Loge, "The smart home concept: our immediate future," 2006 *IEEE International Conference on E-Learning in Industrial Electronics*; (2006), pp. 23–28

[2] H.J. Zainzinge, "An artificial intelligence based tool for home automation using MATLAB," *Tenth IEEE International Conference on Tools with Artificial Intelligence* (1998), pp. 256–261

[3] T. Tajikawa, H. Yoshino, T. Tabaru, S. Shin, "The energy conservation by information appliance," Proceedings *of the 41st SICE Annual Conference*, 5 (2002), pp. 3127–3130

[4] K.S. Rama Rao, Y.T. Meng, S. Taib, M. Syafrudin, "PC based energy management and control system of a building'" *National Power and Energy Conference* (2004), pp. 200–204

[5] A. Gligor, H. Grif, S. Oltean, "Considerations on an intelligent buildings management system for an optimized energy consumption," 2006 *IEEE International Conference on Automation Quality and Testing, Robotics*, 1 (2006), pp. 280–284

[6] F. Baig, A. Mahmood *et al*, "Smart Home Energy Management System for Monitoring and Scheduling of Home Appliances using ZigBee," *J. Basic. Appl. Sci. Res.*, 3(5)880-891, 2013

[7] D.L. Ha, S. Ploix, E. Zamai, Jacomino M, Tabu, "Search for the optimization of household energy consumption," 2006 *IEEE International Conference on Information Reuse and Integration* (2006), pp. 86–92

[8] D.L. Ha, F.F. de Lamotte, Q.H. Huynh, "Real-time dynamic multilevel optimization for demand-side load management," 2007 *IEEE International Conference on Industrial Engineering and Engineering Management* (2007), pp. 945–949
.

 *Corresponding Author: email: nadeemjavaid@comsats.edu.pk, nadeemjavaid@yahoo.com, web: www.njavaid.com